\documentclass[useAMS, usenatbib]{mn2e}
\usepackage{aas_macros}
\usepackage{amsmath}
\usepackage{amssymb} 
\usepackage{graphicx}
\usepackage{subcaption} 
\usepackage{float}
\usepackage{booktabs} 
\usepackage{multirow} 
\usepackage{amsfonts}
\usepackage{lmodern}
\usepackage{bm}

\newcommand{\bc}{\begin{center}}
\newcommand{\ec}{\end{center}}

\let\oldhat\hat
\renewcommand{\hat}[1]{\oldhat{\mathbf{#1}}}
\let\oldbullet\bullet \renewcommand{\bullet}[1][0pt]{%
\mathrel{\raisebox{#1}{$\oldbullet$}}%
}

\title[Quasar outflow in a disc galaxy at $z \sim 5$]{Powerful quasar outflow in a massive disc galaxy at $z \sim 5$} \author[Curtis \& Sijacki]{Michael Curtis$^1$\footnotemark[1], Debora
  Sijacki$^{1}$ \\
  $^1$ Institute of Astronomy and Kavli Institute for Cosmology,
  University of Cambridge, Madingley Road, Cambridge CB$3$ $0$HA, UK}

\begin{document}

\maketitle

\begin{abstract}
There is growing observational evidence of high-redshift quasars launching
energetic, fast outflows, but the effects that these have on their host
galaxies is poorly understood. We employ the moving-mesh code {\small AREPO}
to study the feedback from a quasar that has grown to $\sim 10^9 \mathrm{M}_\odot$ by
$z \sim 5$ and the impact that this has on its host galaxy. Our simulations
use a super-Lagrangian refinement technique to increase the accuracy with
which the interface of the quasar-driven wind and the surrounding gas is
resolved. We find that the feedback injected in these simulations is less
efficient at removing gas from the galaxy than in an identical simulation
  with no super-Lagrangian refinement. This leads to the growth of a massive,
rotationally supported, star-forming disc, co-existing with a powerful
quasar-driven outflow. The properties of our host galaxy,
including the kinematical structure of the gaseous disc and of the outflow,
are in good agreement with current observations. Upcoming ALMA and JWST
observations will be an excellent test of our model and will provide further
clues as to the variance in properties of high-redshift quasar hosts. 
\end{abstract}


\section{Introduction}
\renewcommand{\thefootnote}{\fnsymbol{footnote}}
\footnotetext[1]{E-mail: mc636@ast.cam.ac.uk}

High redshift quasars, thought to be powered by accretion processes on to
supermassive black holes \citep{LyndenBell:69}, are unique probes not only of
black hole growth in the early Universe but also of galaxy formation in extreme
environments. Assembly of black holes with masses in excess of $10^9 \mathrm{M}_{\rm \odot}$
 in less than $1\, {\rm Gyr}$ of cosmic time \citep{Mortlock:11} requires
 copious amounts of low angular momentum gas to be channelled into the innermost regions of a galaxy
 to sustain high accretion rates comparable to or even in excess of the
 Eddington limit. Some of the most promising sites for this physical process
 to take place are very massive proto-clusters embedded within collapsing
 large-scale over-densities and surrounded by a rich web of filaments \citep[e.g.][]{Springel:05N, Sijacki:09, DiMatteo:12,
   Dubois:12b, Costa:14}. Even if a very small fraction of the quasar
 luminosity couples to the surrounding matter this can lead to powerful
 outflows, which have indeed been recently observed in distant quasars \citep[e.g.][]{Maiolino:12,
   Cicone:15}.
   
   A crucial question that then arises is how these outflows
 interact with the host galaxy and affect its gas content, star formation
 rate and morphology. The quasar-driven outflows are sufficiently energetic that
  they may heat and unbind a large fraction of the gas in a galaxy
 \citep{SilkRees, Fabian:99b, King:05}, ridding the galaxy of its star formation
 reservoir 
 and leading to the shutting off of star formation and quenching. On the
 other hand, it is possible that the outflows launched by the central engine cannot
 effectively couple with the surrounding dense interstellar medium (ISM) and
 instead leave the host galaxy without causing major disruption to the ISM \citep{Debuhr:12,
   Bourne:14, Roos:15} or they may even lead to shock-induced star formation bursts
 \citep{Silk:13}. For a given strength of quasar 
 outflow clearly these scenarios can lead to very different host
 galaxy properties. In this Letter we present very high resolution zoom-in
 simulations of a high redshift quasar grown self-consistently in a
 cosmological simulation of a massive proto-cluster region. We employ a recently
 developed super-Lagrangian refinement technique \citep{Curtis:15} to resolve
 the interface between the quasar-launched outflow and its host galaxy with 
 higher spatial resolution, allowing us to study the
 morphological and kinematical imprints that can be directly compared with
 upcoming ALMA and JWST observations.   

\begin{table*}
\bc
\begin{tabular}{ccccccccc}
\hline
\bf{Simulation type} & $M_{200}$ ($\mathrm{M}_{\rm \odot}$) & $M_\mathrm{DM,200}$
   ($\mathrm{M}_{\rm \odot}$) & $M_\mathrm{star,200}$ ($\mathrm{M}_{\rm \odot}$) & $M_\mathrm{gas,200}$
   ($\mathrm{M}_{\rm \odot}$) & $R_{200}$ (kpc) & SFR ($\mathrm{M}_{\rm \odot}\, {\rm yr}^{-1}$)\\ \hline
Default & $8.04 \times 10^{12}$ & $6.77 \times 10^{12}$ & $4.07 \times 10^{11}$ & $9.15 \times 10^{11}$ & 108 & 96.5 \\
Refinement & $8.11 \times 10^{12}$ & $6.70 \times 10^{12}$ & $4.55 \times 10^{11}$ & $8.48 \times 10^{11}$ & 108 & 397 \\
\hline
\end{tabular}
\caption{Properties of the halo hosting the largest supermassive black hole at
  $z=4.9$. We list the total virial mass, as well as the mass in dark matter, stars
  and gas together with the virial radius and the star formation rate integrated over the whole halo.} 
\label{tb_galaxy_properties} 
\ec
\end{table*}

\section{Methodology}

We use the finite volume moving mesh code {\small AREPO}
\citep{Springel:10} which adopts the TreePM approach for gravity and a Voronoi
mesh to discretise the fluid. We select one of the most massive
haloes from the Millennium simulation \citep{Springel:05N} and resimulate it
at high resolution, including baryons as in \citet{Sijacki:09,
  Costa:14}. The spatial and mass resolutions of our simulation are: $\epsilon_{\rm
  grav} = 1 \,{\rm kpc}$ (comoving), $m_{\rm DM} = 8 \times 10^6 \mathrm{M}_{\rm
  \odot}$ and $m_{\rm gas} = 10^6 \mathrm{M}_{\rm \odot}$, 
where $m_{\rm gas}$ is  
the mean cell mass at the beginning of the simulation. We include primordial
gas cooling and heating and a UV background as in \citet{Sijacki:09} and a
sub-grid model for star formation and associated feedback \citep{SF}. No
galactic outflows are explicitly included. We seed $10^5 \mathrm{M}_{\rm \odot} \,
h^{-1}$ black holes in haloes with $10^{10} \mathrm{M}_{\rm \odot} \, h^{-1}$ which then
grow at the Bondi rate, capped at the Eddington limit, and through black hole mergers. Feedback consists of
thermal energy which is injected into the surrounding gas cells within the
  black hole smoothing length and weighted by mass. In both simulations
  this region contains $64$ times the mean gas cell mass.
For full details of the black hole model see \citet{Springel:05, Sijacki:09, 
  Curtis:15}.  

We present the results of two otherwise identical simulations, one of
which adopts the super-Lagrangian refinement 
method detailed in \citet{Curtis:15}, which allows us to increase the
resolution of our simulations around the black holes. To do
this, we split and merge cells over an adaptive region defined by the
smoothing length of each black hole (which is typically around $1 \, {\rm kpc}$ for the largest black hole), forcing the innermost cells to be the size of the Bondi radius of each black hole.  For $z \lesssim 7$ when the black hole growth reaches a plateau, the mean cell size in the refined
region is $65 \, {\rm pc}$, while the smallest cells have a size of $1 \, {\rm
pc}$ and a mass of $\sim 0.1 \,\mathrm{M}_{\rm \odot}$. The region of refinement itself is typically $\sim 1 \, {\rm kpc}$ (physical). Our aim is to examine how a better resolution in the region in which feedback is injected, without otherwise changing the parameters of the feedback, can make a difference in our understanding of the environment and host galaxy properties of high redshift quasars. The results presented below, where not otherwise indicated, are from our simulation with the refinement scheme and all properties are given in physical units.

\begin{figure*}
\centering
\includegraphics[width=0.42\textwidth]{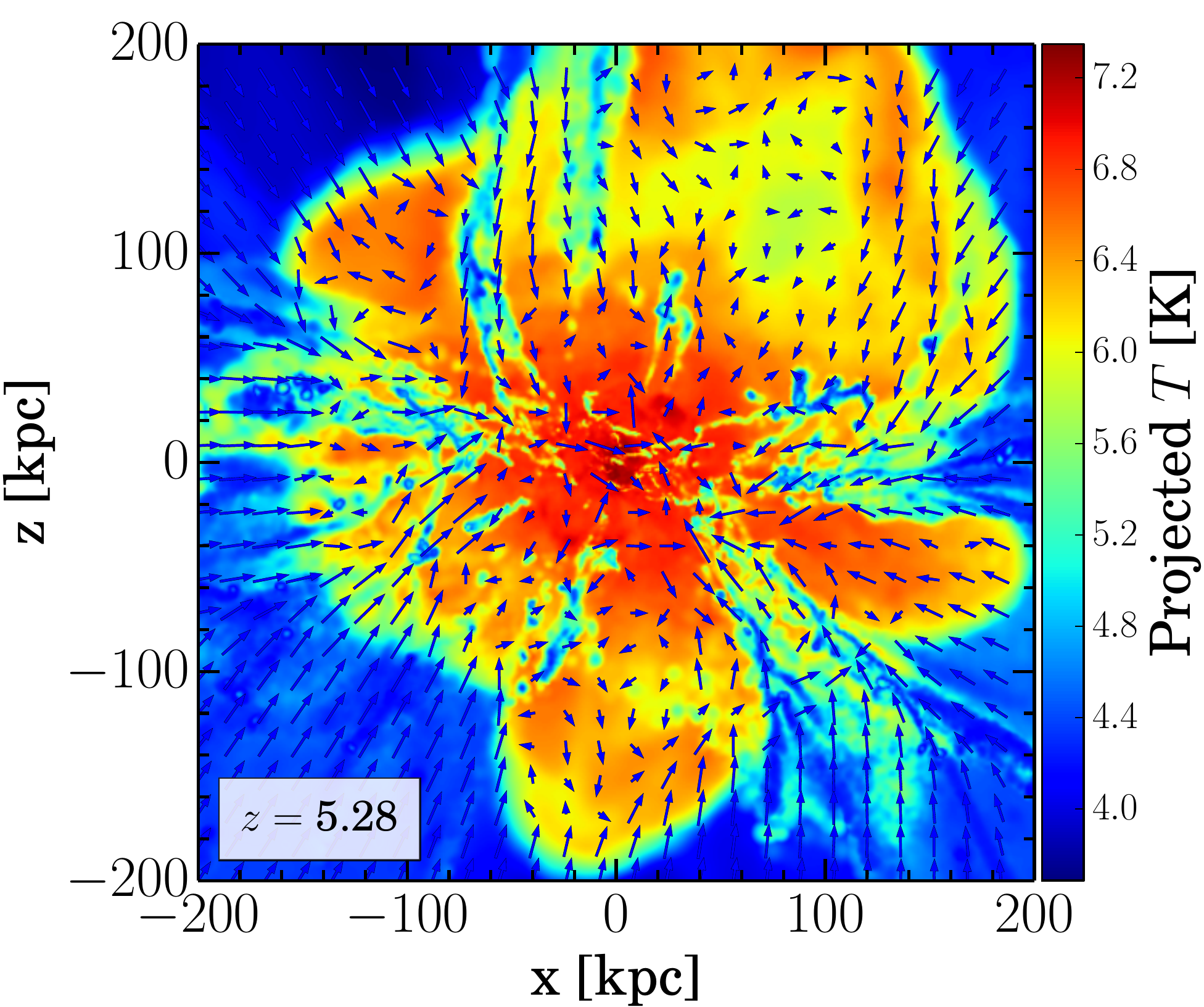}
\includegraphics[width=0.42\textwidth]{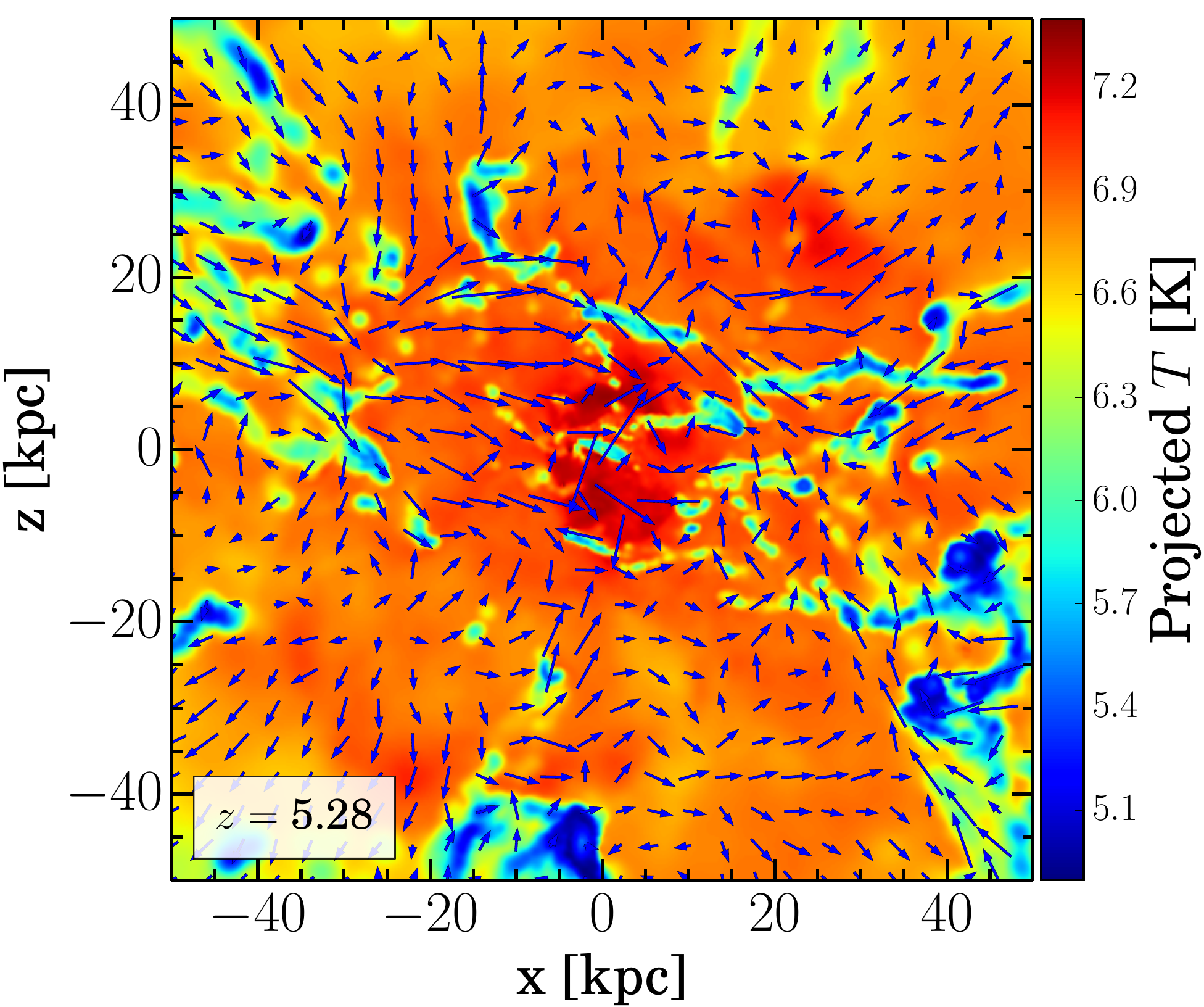}
\includegraphics[width=0.42\textwidth]{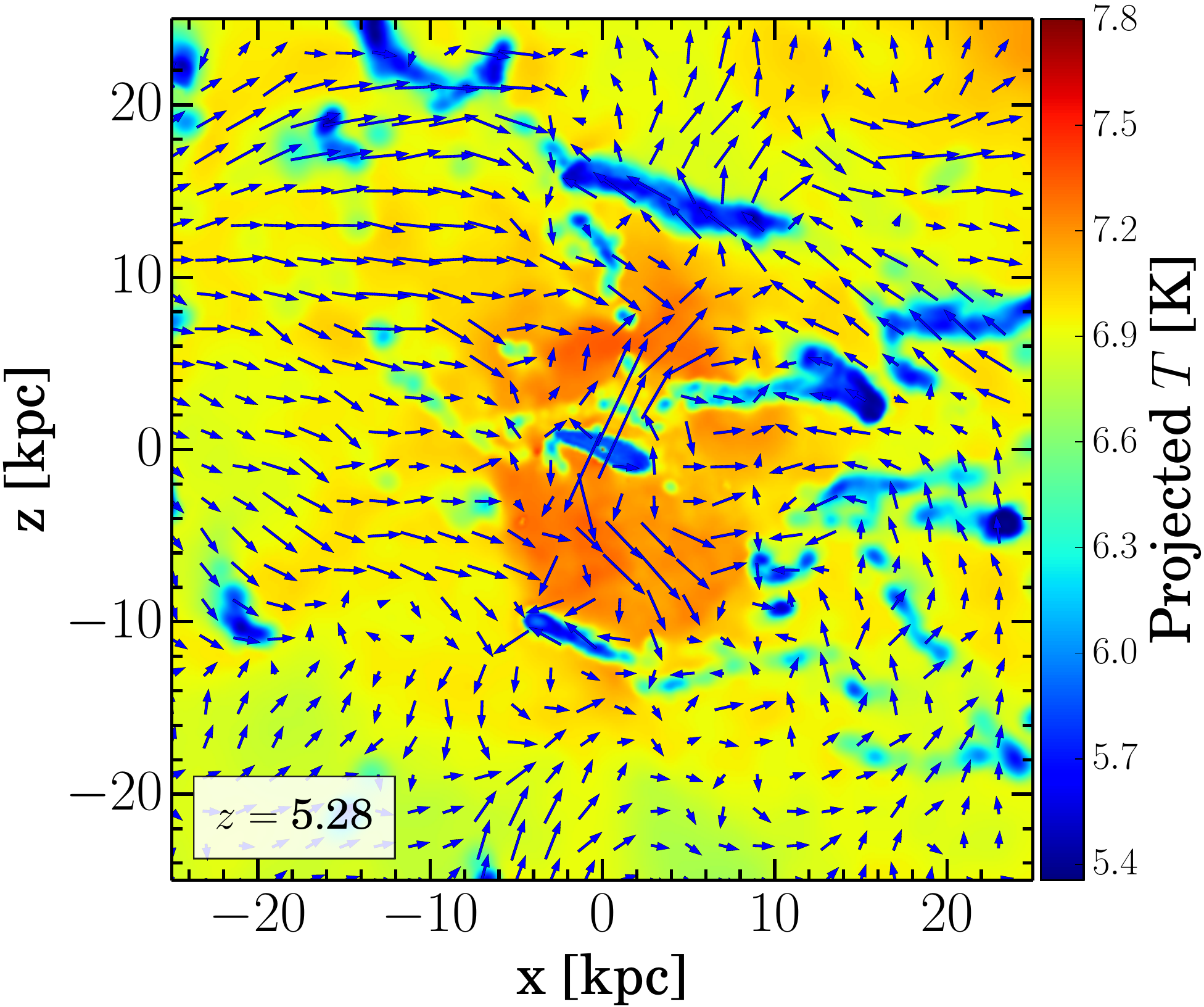}
\includegraphics[width=0.42\textwidth]{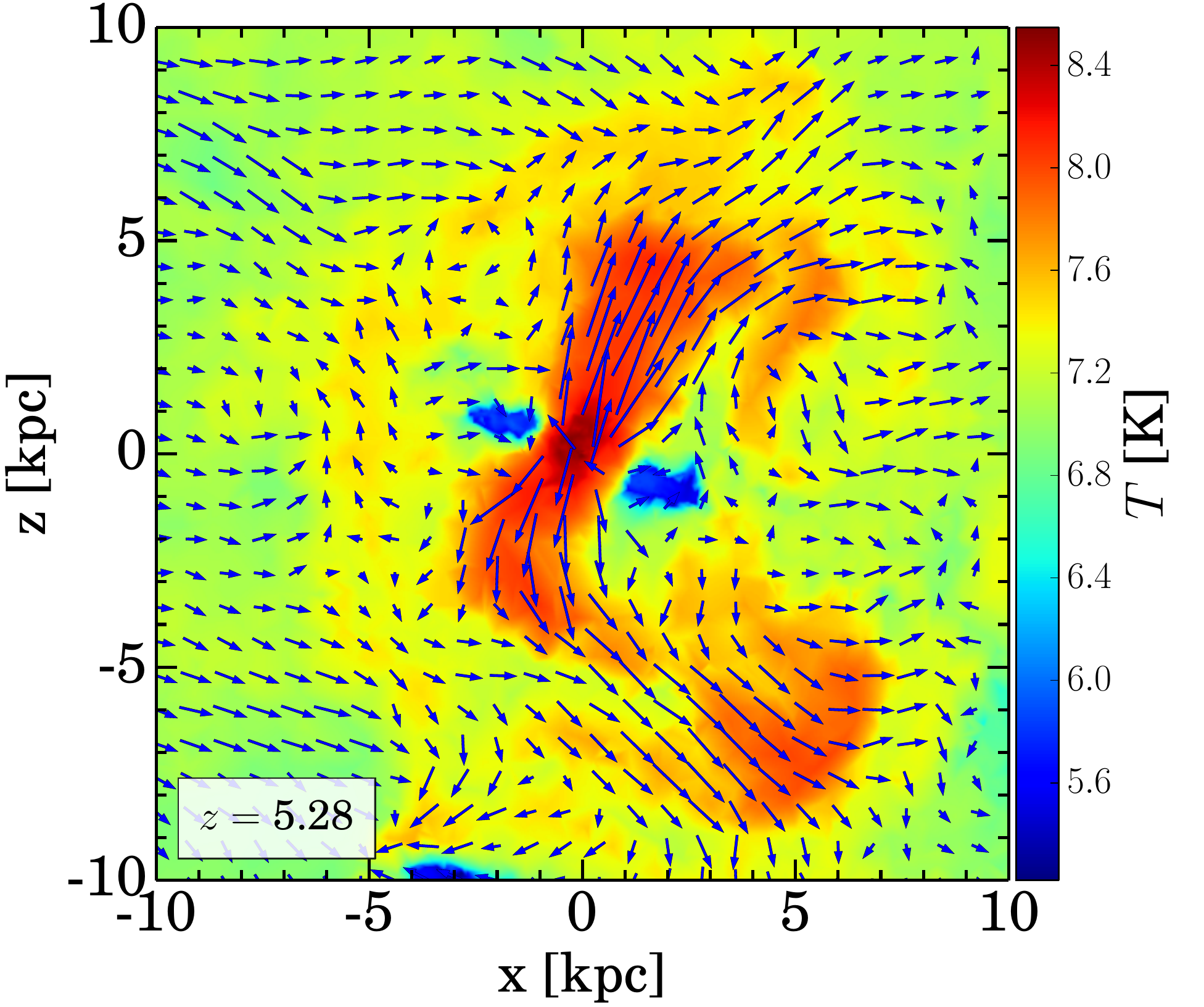}
\caption{The temperature distribution of the gas surrounding the
  largest supermassive black hole at $z=5.3$. In the top row we show the
  projected temperature, mass-averaged over a slice $50\, {\rm kpc}$ thick,
  whilst in 
  the bottom left panel the projection is averaged over $25 \, {\rm kpc}$. The
  velocity field of the gas is over-plotted with arrows, with the mean size of
  the 
  arrows representing a velocity of $250 \, {\rm km \, s^{-1}}$. In the bottom
  right-hand panel 
  we show the gas temperature of a slice 
  through the centre of the galaxy. Here we can see a cold gaseous disc
  whose net angular momentum axis is aligned with that of a hot,
  fast-moving bipolar outflow, launched by the
  central black hole.}   
\label{fg_proj_temp}
\end{figure*}

\section{Results}

\subsection{The large scale distribution}

After the first black holes are seeded at $z \sim 15$, a subset of simulated black holes
begin growing rapidly for $z \lesssim 10$, quickly reaching the Eddington limit.
 In this phase, the feedback is insufficient to prevent gas inflows
on to massive haloes and so does little to affect the black hole accretion
\citep[see also][]{Sijacki:09, Costa:14} or the star formation, which reaches a 
peak of $1200 \mathrm{M}_{\rm \odot} \, {\rm yr^{-1}}$ at $z = 7$. The black hole growth and host galaxy
properties in the simulations with and without the super-Lagrangian refinement are very
similar during this phase. The differences begin to occur during the end of
the Eddington growth. At this point, the black hole has reached a sufficiently
high mass to cut off its own accretion flow, which for the most massive black hole in our
 simulations happens at $z \sim 7$. We find that the nature of
 this initial shut off and the manner in which the feedback continues to
 regulate the black hole growth after this phase is important in determining
 the galaxy properties, and we focus our subsequent analysis on
 this later period. It is worth noting that in the simulation with
 super-Lagrangian refinement the black hole growth is somewhat more efficient for
 $z \le 7$  and the final mass of the most massive black hole is $1.2$ times
 higher at $z \sim 5$, reaching $1.8 \times 10^9 \mathrm{M}_{\rm \odot}$. The bolometric luminosity of the 
 quasar in this period varies between $10^{44}$ and $10^{46} {\rm erg \, s^{-1}}$.

In Fig.~\ref{fg_proj_temp} we show gas temperature maps centred on the
largest halo in our simulation (detailed halo properties are
listed in Table~\ref{tb_galaxy_properties}). Here, the rich web of cold
filaments is feeding the halo with gas from large scales, following the
distribution of the dark matter. As the filaments reach the virial radius the surrounding gas
rises in temperature, from $ \sim 10^4\, {\rm K}$ to 
much higher temperatures of around a few $10^6\, {\rm K}$. This is mostly the
result of the gas virially shocking, in line with the classical analytical
models of galaxy formation \citep{Rees:77, Silk:77,
  WhiteFrenk:91}, whereby the cooling radius of the in-falling gas is much
smaller than the virial radius, allowing for a quasi-hydrostatic atmosphere to
form in the halo. In addition to this, feedback from the central black hole
heats 
the gas to a high temperature of a few $10^8 \, {\rm K}$. In line with
previous simulations \citep[see e.g.][]{Keres:05, Ocvirk:08, Nelson:13} the
filaments are not totally disrupted at the virial radius and the cold filamentary accretion
continues down to smaller scales \citep{DiMatteo:12, Dubois:12b,
  Costa:14}. While the filaments are largely disrupted roughly half-way
through the halo, the residual accretion of the cold gas persists across
the successively smaller scales shown in Fig.~\ref{fg_proj_temp},
connecting large scale filaments with the central galaxy at the bottom of the potential well. 

The bottom right-hand panel of Fig.~\ref{fg_proj_temp} shows the gas temperature of
a slice through the centre of the halo. The gas that has reached this far into
the halo has 
circularized into a massive cold gaseous disc, which forms not long after the end of the Eddington limited phase of accretion, and is in place by $z \sim 6$. Detailed galaxy properties are listed in
Table~\ref{tb_disc_properties}. Note that the axis of rotation of the disc is largely
perpendicular to the orientation of the main filaments. Parallel to the rotational axis of the disc is the hot, outflowing gas that has been launched by the powerful feedback from the central
supermassive black hole. This is entirely hydrodynamically driven - the black
hole injects thermal energy into the gas in the very central region which then
expands and rises in the comparatively cool medium of the disc and surrounding
warm gas. The peak velocity of the outflow, $2800 \, {\rm km\, s}^{-1}$,
occurs at $5 \, {\rm kpc}$ from the black hole, as the centre of the outflow
accelerates slightly, before dropping off. The mass averaged velocity of the
outflowing gas, however, is fastest in the centre of the outflow and at its
base where the pressure gradient is strongest. 

The cold gaseous disc shown here is not present in our
simulation without super-Lagrangian refinement - a similar structure forms at
the same time, i.e. at $z \sim 7$, but in the case without refinement, the cold
gas is completely 
overwhelmed by the feedback from the black hole. This is particularly
interesting given that, in the refinement simulation, the black hole grows
slightly faster and, as such, more cumulative feedback energy has been
injected into the gas by $z \sim 5$. This underlines
the importance of resolving the inner region of the outflowing gas - if the
resolution is insufficient (both spatially or temporally) then the feedback
will blow away the cold gas before the hot quasar-driven wind is able to rise
out of the galaxy and subsequent predictions about the morphology of the host galaxy will likely be incorrect.

\subsection{Properties of the inflow and outflow}

\begin{figure*}
\centering
\includegraphics[width=0.4\textwidth]{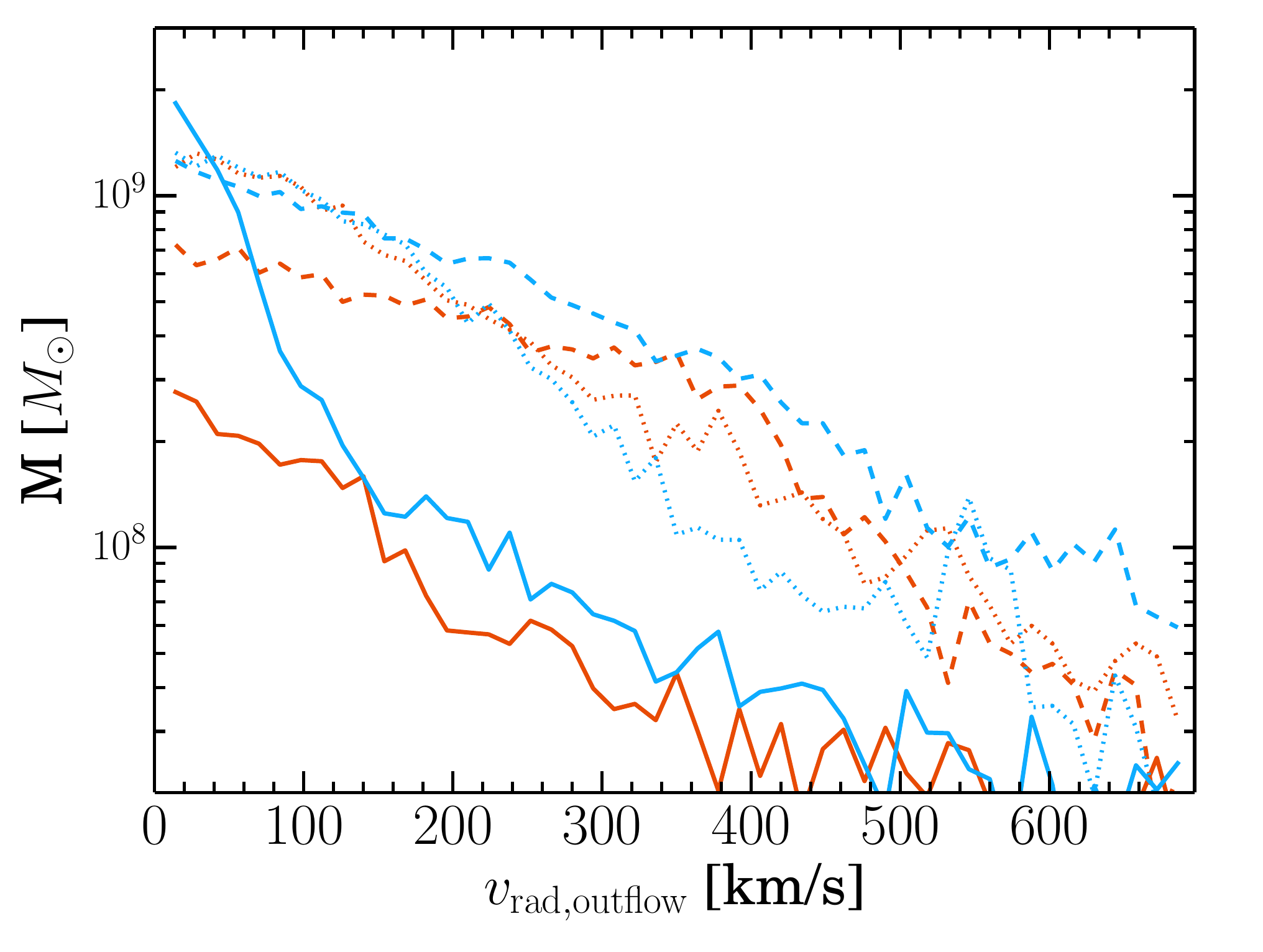}
\includegraphics[width=0.4\textwidth]{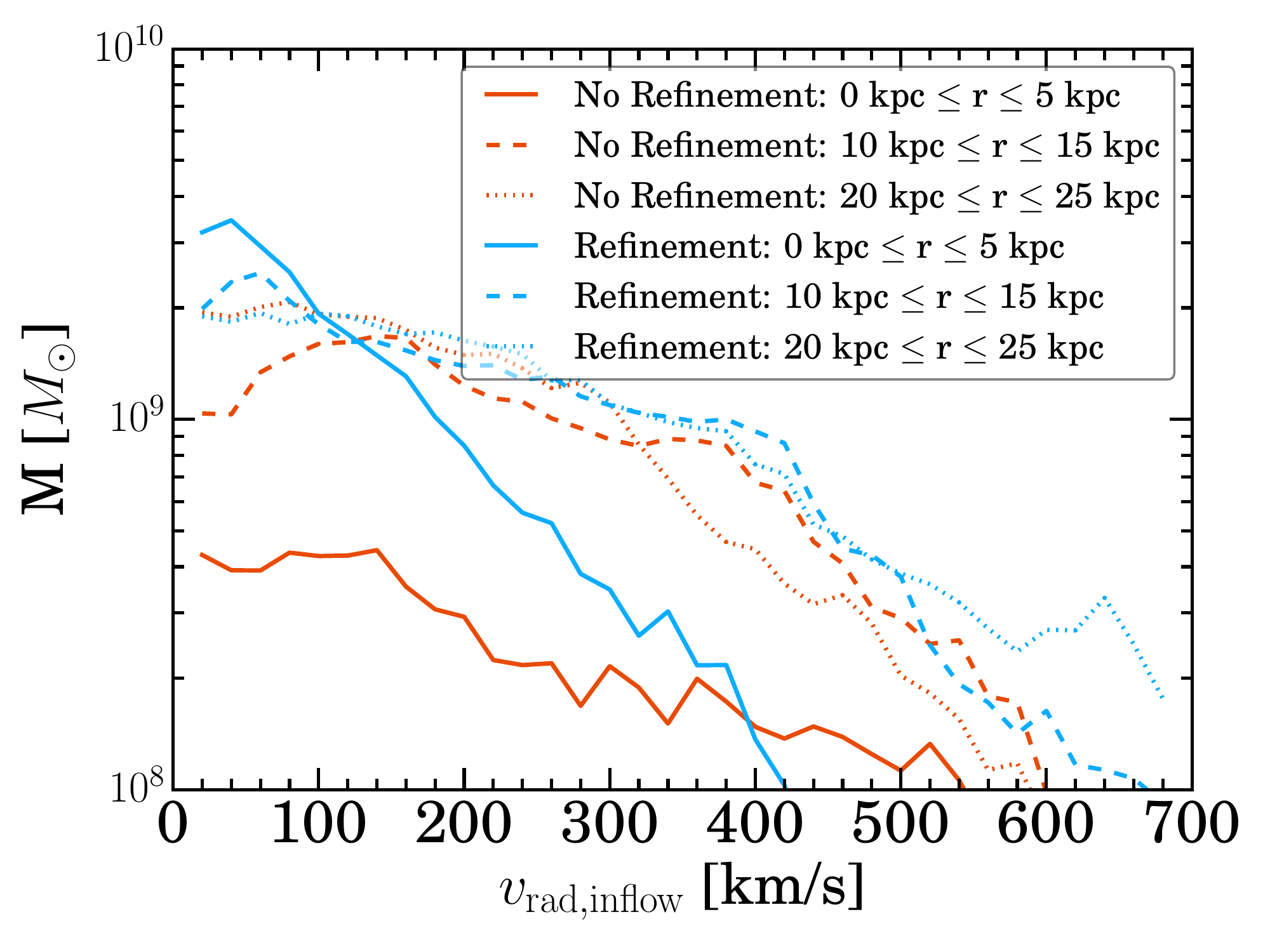}
\caption{The velocity distribution of the gas for
  simulations with refinement 
  (blue) and without (red) at $z = 5.3$. In the left-hand panel, we show the mass of outflowing gas as a function of speed while in the right-hand panel, 
  we plot the mass of gas with inflowing radial orbits. In both, we show the 
  relevant quantity summed over three spherical shells.}  
\label{fg_inflow_outflow}
\end{figure*}

In Fig.~\ref{fg_inflow_outflow}, we show the gas mass distribution as a function of
its velocity for both simulations. Our intent here is
to investigate what difference the use of refinement has on the kinematic
properties of the gas, even at distances much larger than those at which we
increase the resolution, i.e. at $\sim 1 \, {\rm kpc}$. These plots are for
our simulations at $z=4.9$, but the results are very similar from $z = 7.0$
onwards, when the black hole is in its post Eddington regime.  

In the left-hand panel of Fig.~\ref{fg_inflow_outflow}, we show the velocity
distribution of the outflowing
gas. At all radii, the two simulations agree reasonably well,
predicting a similar gas mass distribution, but overall there is more
of the outflowing gas in the simulation with refinement. We
have also estimated the total mass in the outflow, which is $9 \times 10^{10}
\mathrm{M}_{\rm \odot}$ within $25 {\rm kpc}$, its momentum flux is $5 -
40 \, {\rm Lc^{-1}}$, while the outflow rate is
$\sim 700 \,{\rm \mathrm{M}_{\rm \odot} \, yr ^{-1}}$ (if we consider speeds higher than
$400 \,{\rm km \, s^{-1}}$ this number drops to $\sim 300 \,{\rm \mathrm{M}_{\rm \odot} \, yr ^{-1}}$).

In the right-hand panel of Fig.~\ref{fg_inflow_outflow} we show a similar plot,
but for the inflowing gas. There is a significant difference in the
mass distribution at low radii - the non-refinement simulation exhibits little
inflow across all velocity bins, whilst the refinement simulation shows a much
larger mass of inflowing gas at lower velocities. This difference is still
sizeable at $10\, {\rm kpc}$ out from the black hole, an order of magnitude
outside of our refinement region. By the time we reach $20\, {\rm kpc}$, the
two simulations agree well. The difference, in both cases, is a
  reflection of the simulation resolution 
  and hence the spatial distribution of the outflow. With
  refinement, the outflow is tightly collimated with a small opening angle (as the
  cells are not enforced to maintain approximately constant mass), and the gas
  rises vertically out of the galaxy without interacting much with the cold
  disc of accreting gas. Without refinement the hot gas expands, leading to a
  very coarse effective resolution and it interacts with the inflowing gas,
  reducing both the inflow and outflow at small radii. 

\subsection{Galaxy morphology}

\begin{table}
\bc
\begin{tabular}{ccccc}
\hline
\bf{Disc} & $R_\mathrm{visual}$ & $R_{\mathrm{mass}/2}$
& $M(< R_\mathrm{visual})$ &
$\mathrm{SFR}(<R_\mathrm{visual})$ \\
 & (kpc) & (kpc) & ($ 10^{10} \mathrm{M}_{\rm \odot}$) & ($\mathrm{M}_{\rm \odot}\, {\rm yr}^{-1}$) \\\hline
Gas       & 3.5 & 2.05 & 3.53 & 332 \\
Stellar   & 3.5 & 0.86 & 27.5 &  - \\
\hline
\end{tabular}
\caption{Galaxy properties at $z=4.9$ listing gas and stellar disc visual
  radius,  half-mass radius, mass and star formation rate within the visual
  radius, respectively.}
\label{tb_disc_properties} 
\ec
\end{table}

\label{morph}

\begin{figure*}
\centering
\includegraphics[width=0.4\textwidth]{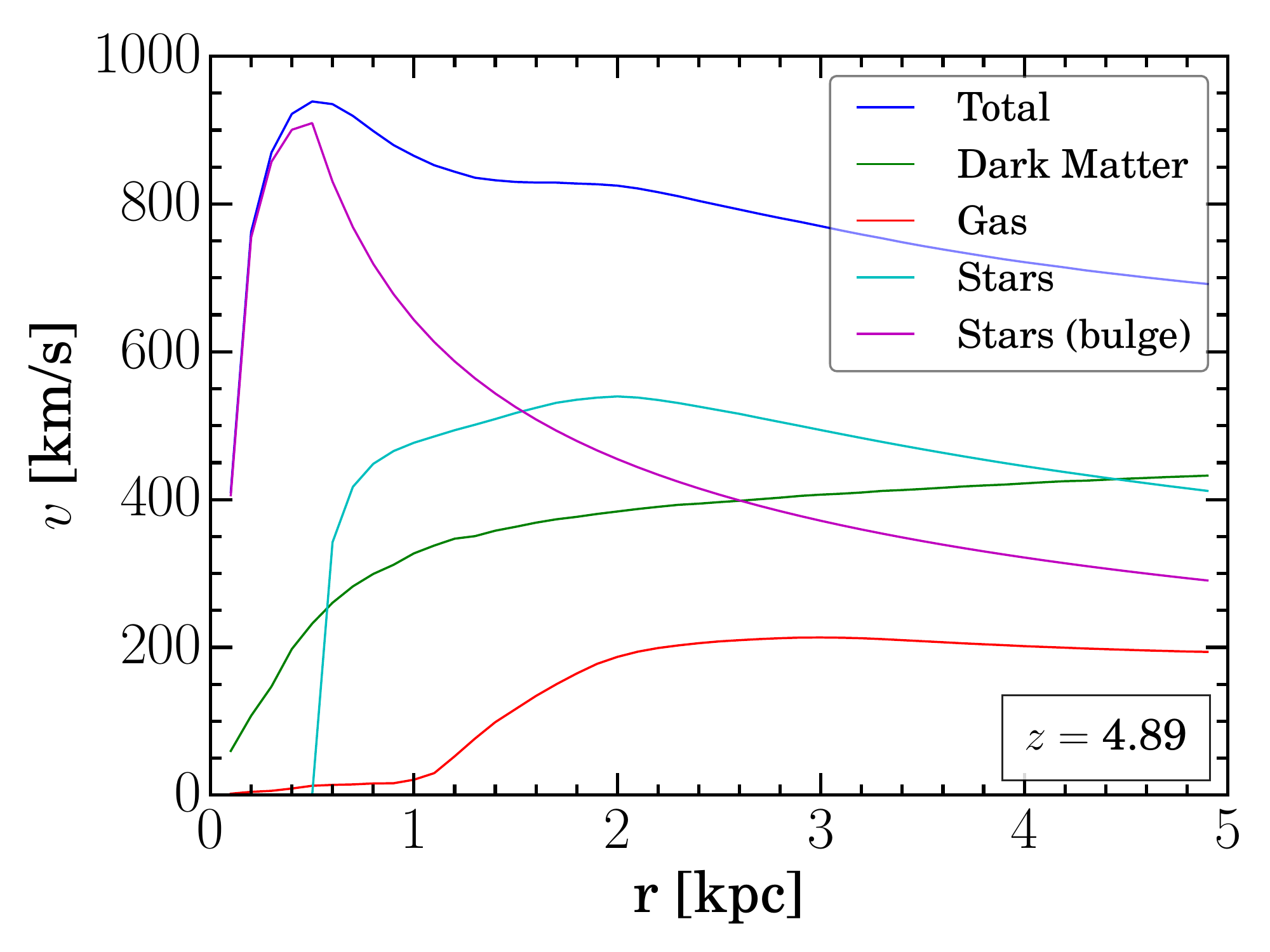}
\includegraphics[width=0.4\textwidth]{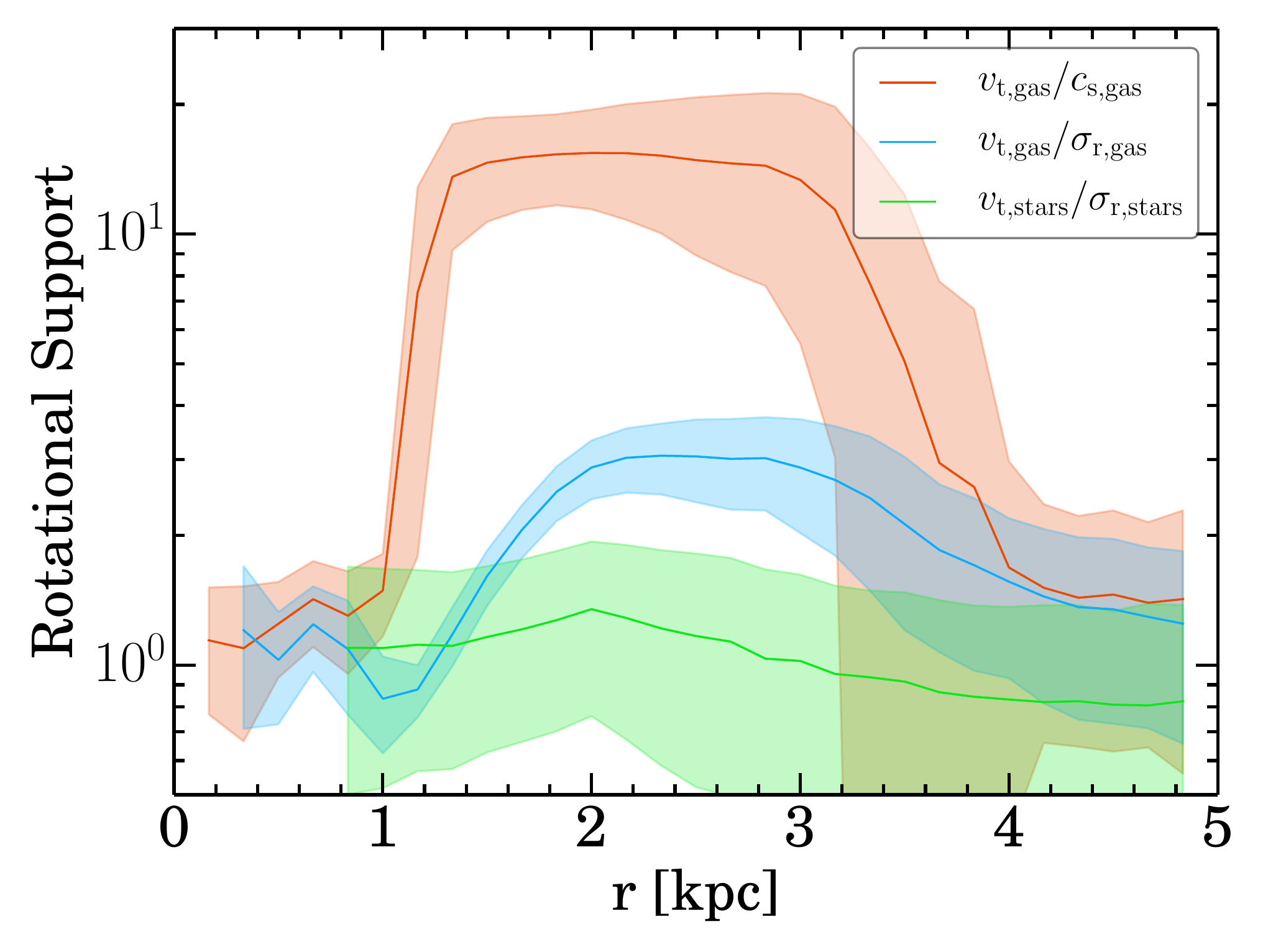}
\caption{Left: rotation curves of different components, based on the enclosed mass. Right: the tangential speed of the gas divided by the sound speed
  (red), and by the enclosed radial velocity dispersion of the
  gas (blue). The tangential
  speed of the stars divided by the enclosed radial velocity dispersion of the
  stars is shown in green. Lines denote the mean and the shaded area is $1 \,
  \sigma$ scatter.}    
\label{fg_rot_sup}
\end{figure*}

\begin{figure*}
\centering
\includegraphics[width=0.42\textwidth]{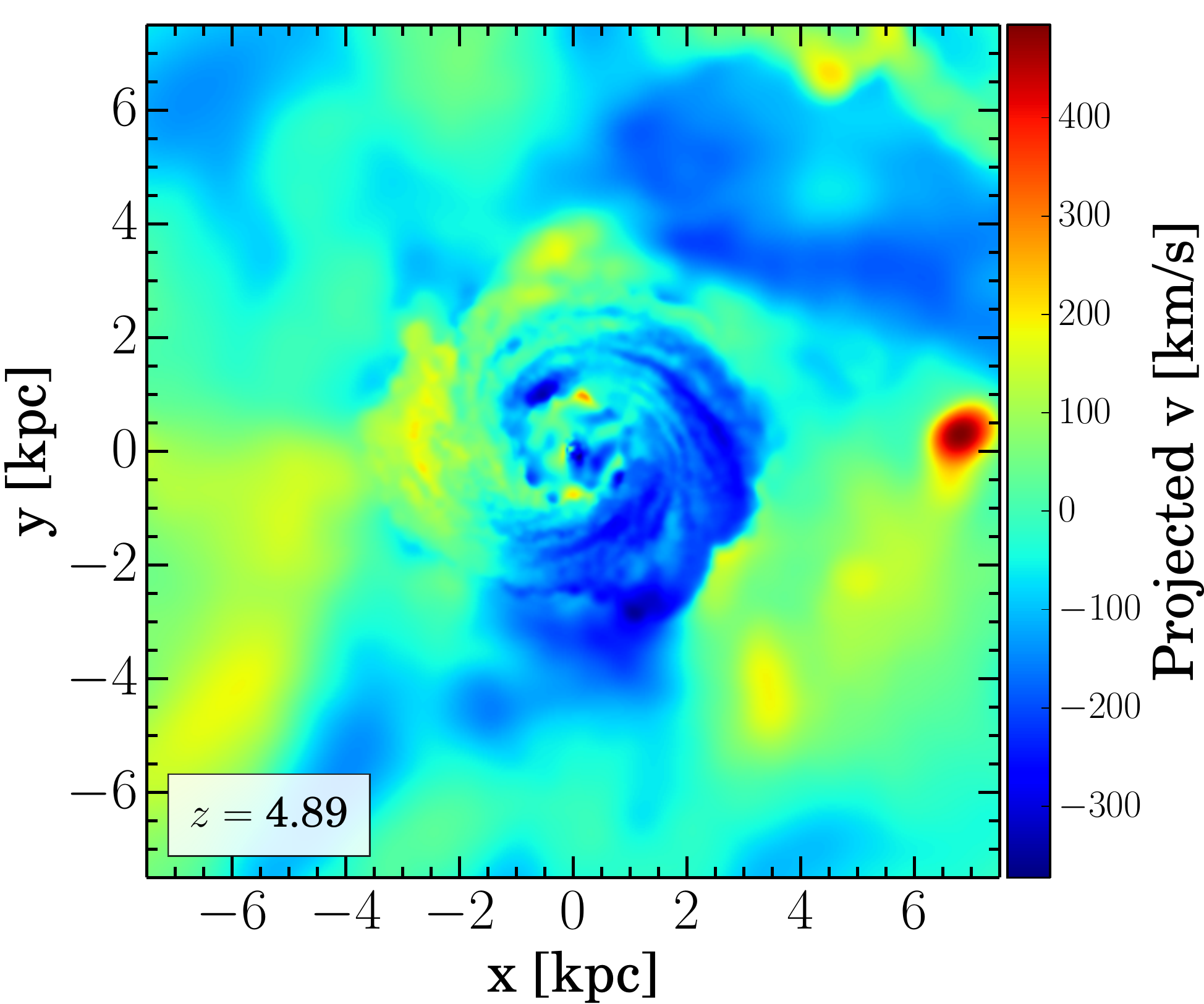}
\includegraphics[width=0.42\textwidth]{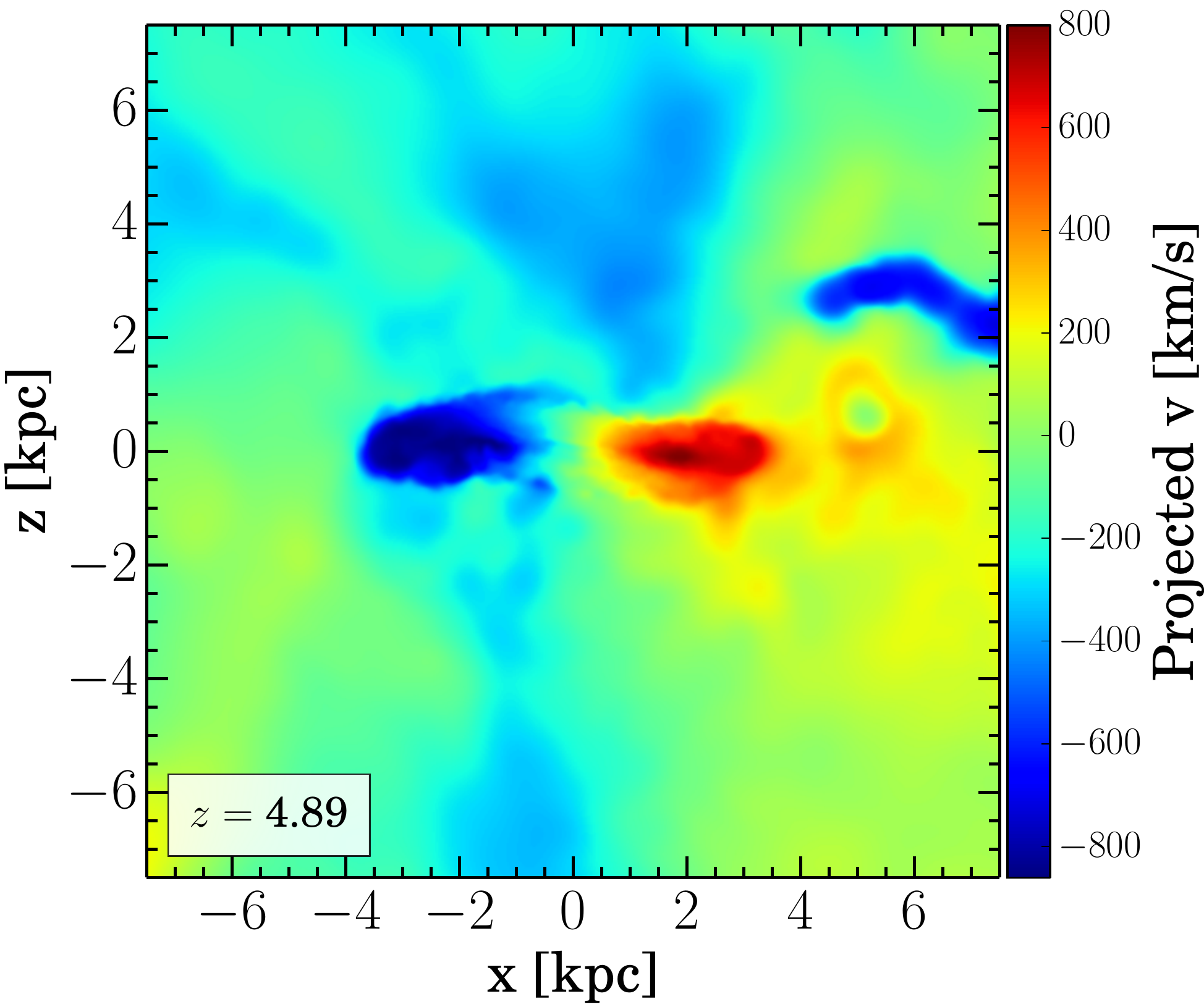}\\
\caption{Gas velocity maps of the central galaxy in a face-on (left)
  and edge-on (right) projection. The projection depth is $13 \, {\rm
    kpc}$. While the rotational signature is weak for the face-on view, it is
  clearly detectable for the edge-on view with speeds up to $\sim 700 \, {\rm km \,
    s^{-1}}$.}    
\label{fg_galaxy_velocity}
\end{figure*}

We now focus on the morphological properties of the quasar host galaxy in our
simulation with refinement. The black hole sits at the centre of an 
exponential stellar disc, with a prominent and kinematically distinct central
bulge component. The  
$M_\mathrm{BH}-M_\mathrm{bulge}$ ratio for our galaxy is $0.02$, which is within the scatter of the
observed relation of \citet{Kormendy:13}. The abundance matching constraints 
of \citet{Moster} suggest that a dark matter halo of mass $6.8 \times 10^{12} \mathrm{M}_{\rm \odot}$ at $z
  = 4$ should on average host a central galaxy with a stellar mass of
$\sim 10^{11} \mathrm{M}_{\rm \odot}$, indicating that while we overproduce around twice as
many stars as predicted by the mean relation we are within the 1$\sigma$
  scatter, which has an upper bound of $2.7 \times 10^{11} \mathrm{M}_{\rm \odot}$. We do not include strong stellar feedback in our simulations in
the form of energetic supernova-driven outflows, which may likely bring our simulated
disc stellar mass closer to the mean relation. 

In the left-hand panel of Fig.~\ref{fg_rot_sup} we show the rotation curves of
different components, with stellar bulge dominating in the centre, followed by
the stellar disc at the intermediate range of radii and finally by dark matter
outside of the galaxy. In the right-hand panel we plot the tangential velocity
of the gas divided by the sound speed of 
the gas and the radial velocity dispersion of the gas,
respectively. The tangential velocity of the stars divided by their radial
velocity dispersion is shown as well, indicating much larger velocity
dispersion support. The gas disc is aligned with the stellar component and covers a similar radial
extent. The surface density  
profile is however not well modelled by an exponential, especially in the inner
region. In Fig.~\ref{fg_galaxy_velocity} we show the projected line of sight
velocity of 
the gas centred on the galaxy. There is a clear signature of
the rotating gas disc both when viewed face-on and edge-on with typical
velocities of $200 \, {\rm km \, s^{-1}}$ and $700 \, {\rm km \, s^{-1}}$,
respectively.  

\section{Discussion}

Recent observations have started to shed light on the properties of quasar
feedback at high redshifts and several studies have found evidence of
large-scale outflows \citep[e.g][]{Maiolino:12, Cicone:15}. However, there
are currently very few observed examples which can constrain in detail both
the quasar 
outflow and the host galaxy properties, especially at high redshift. A notable
exception, but in the local Universe, is Mrk 231, a ULIRG galaxy with a $5
\times 10^{45} \, {\rm erg\, s^{-1}}$ quasar at its core. The inferred star
formation rate of the galaxy of $200 {\rm M_{\rm \odot} \, yr ^{-1}}$, its
regular rotation pattern together with the outflow rate of $700 {\rm M_{\rm
    \odot} \, yr ^{-1}}$ and velocity of $750 {\rm km \, s^{-1}}$
\citep{Feruglio:10} agrees very well with our 
results, indicating that our simulated system could be a high 
redshift counterpart of Mrk 231. Recent ALMA observations by
\citet{Carniani:13} have found a quasar - SMG pair at $z \sim 4.7$, where host
galaxies exhibit a rotationally supported geometry with a range of velocities and disc sizes similar to our findings, albeit with no
detected quasar outflow. Moreover, ALMA observations of five luminous quasars at
$z \sim 6$ by \citet{Wang:13} with a range of black hole masses and host
galaxy dynamical masses bracketing our results, find evidence of velocity
gradients indicative of rotationally supported discs. It is worth noting
however that observed quasar host galaxies are more likely to be viewed
face-on, and this selection bias needs to be taken into account when comparing
to our simulation results.

The results presented in this Letter indicate that a more careful treatment of
the black hole 
feedback injection can lead to feedback having less of an impact on galaxy
properties \citep[see also][]{Dubois2015, Feng2015}, at least concerning the
central galaxy over limited periods of 
time. However, the injected energy can have a longer term impact - it will
heat the gas outside of the central galaxy in the circum-galactic and
inter-galactic medium. This will affect the subsequent gas cooling on to the galaxy which, if there is sufficient energy, may ultimately lead to the starvation and quenching of star formation.

In light of this, it will be increasingly important in future work to attempt
to model the ISM, ideally down to parsec scales and below. High density clumps
within the ISM will be more resistant to 
destruction by the central engine, and this will further complicate the
picture of how the feedback energy couples to the surrounding medium and
quenches star formation. Future ALMA and JWST 
observations will be crucial in this regard, as they will allow detailed
comparison with the theoretical models and also provide larger statistical
samples to pin down the morphological variety of galaxies hosting powerful quasar
outflows.

\section*{Acknowledgements}
We thank Roberto Maiolino, Martin Haehnelt and Ewald Puchwein
for many helpful suggestions on the manuscript.
MC is supported by the STFC and DS acknowledges support by the ERC
Starting Grant 638707 ``Black holes and their host galaxies: co-evolution
across cosmic time''.  
This work was performed on: DiRAC Darwin
Supercomputer (University of Cambridge HPCS; Higher Education Funding
Council for England and STFC); DiRAC Complexity
system (University of Leicester IT Services; BIS
National E-Infrastructure grant ST/K000373/1 and
STFC DiRAC grant ST/K0003259/1); the COSMA Data Centric system (Durham
University; BIS National E-infrastructure grant
ST/K00042X/1, STFC grant ST/K00087X/1, DiRAC Operations
grant ST/K003267/1 and Durham University). DiRAC is part of the
National E-Infrastructure.

  \bibliographystyle{mn2e} 
  \bibliography{references}

\end{document}